\documentclass[runningheads]{llncs}
\usepackage{amsmath}
\usepackage[T1]{fontenc}
\usepackage{graphicx}
\begin{document}
\title{Hardware-Efficient FPGA Implementation of Sigmoid Function Using Mixed-Radix Hyperbolic Rotation CORDIC}
\titlerunning{Hardware-Efficient Sigmoid Activation Using MR-HRC}
%
\author{Chintan Panchal\inst{1} \and
Ankur Changela\inst{1} \and
Mohendra Roy\inst{1}}
\authorrunning{P. Chintan et al.}
%
\institute{Pandit Deendayal Energy University, Gujarat, India\\
\email{Corresponding authors: Ankur.Changela@sot.pdpu.ac.in and mohendra.roy@ieee.org}}
\maketitle              
\begin{abstract}
Efficient hardware implementation of nonlinear activation functions is a crucial task in deploying artificial neural networks on resource-constrained and edge devices such as Field-Programmable Gate Arrays (FPGAs). The sigmoid activation function is more popular among researchers for probabilistic output, binary classification, and gating mechanisms in recurrent neural networks, despite its exponential computations. This paper presents a hardware-efficient FPGA implementation of the sigmoid activation function using a mixed-radix CORDIC-based architecture. The proposed approach uses the mathematical relationship between the sigmoid and hyperbolic tangent functions. The input range is normalized to $\pm1$, enabling the corresponding Tanh computation to operate within a reduced range of $\pm0.5$, which significantly improves convergence behavior.
To achieve high accuracy with minimal hardware overhead, a modified mixed-radix hyperbolic rotation CORDIC (MR-HRC) algorithm combining radix-2 and radix-4 iterations is presented. The initial radix-2 stage ensures stable convergence, while the subsequent radix-4 stage accelerates convergence without the need for scale-factor compensation. Later, the radix-2 linear vectoring CORDIC (R2-LVC) stage is used to compute the hyperbolic tangent by dividing hyperbolic sine and cosine derived using the MR-HRC algorithm. The entire architecture is fully pipelined and implemented on an FPGA. The proposed design is implemented on an Xilinx Virtex-7 FPGA using a 16-bit fixed-point representation. Experimental results demonstrate a significant reduction in hardware utilization, achieving only 835 logic slices with zero DSP usage. Furthermore, the design attains a mean absolute error of $4.23 \times 10^{-4}$, showing better performance compared to recent sigmoid implementations. 

\end{abstract}
\section{Introduction}

Neuroengineering has made great strides over the past few years, largely due to increasing interest in smaller, faster, cheaper means of providing ANNs through integrated circuits (IC). Neuromorphic hardware is designed to operate similarly to the biological neural networks in the way they process information and produce output. This type of system provides a means to implement ANNs across a much broader application space, including: high-bandwidth video applications, cryptography, IoT, robotics, and telecommunications, where speed, size, and energy efficiency are of utmost importance. 

Field-Programmable Gate Arrays (FPGAs) are well-suited for these applications due to their ability to support highly parallel computing architectures, shorter development cycles, and lower production costs for medium-scale projects.  FPGAs offer advantages over traditional Central Processing Units (CPUs) and Graphics Processing Units (GPUs) in terms of energy efficiency and compact design.  However, implementing ANNs directly on FPGAs presents unique challenges, particularly in the computation of neuron activation functions.  These challenges stem from the need to balance hardware resource constraints with numerical precision, especially when using fixed-point arithmetic instead of floating-point calculations. 

\begin{figure}
    \centering
    \includegraphics[width=0.5\linewidth]{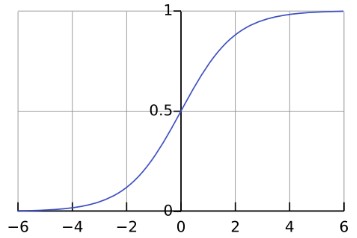}
    \caption{Sigmoid activation function over a range $[-6,6]$}
    \label{fig0}
\end{figure}

Among the various activation functions used in neural networks, the sigmoid function, as shown in Fig. \ref{fig0} is one of the most widely employed due to its nonlinear properties and ability to map input values to a range between 0 and 1.  It plays a crucial role in shallow networks like Multilayer Perceptrons (MLPs) during both inference and training phases, as well as in recurrent neural networks such as Long Short-Term Memory (LSTM) networks and Gated Recurrent Units (GRUs).  While Rectified Linear Units (ReLU) and their variants have gained popularity in deep learning applications, the sigmoid function remains essential for tasks involving binary decision-making, probabilistic outputs, and structured attention mechanisms.

Despite its importance, directly implementing the sigmoid function in hardware is computationally expensive due to the complex operations involved, such as exponentiation and division.  This paper addresses this challenge by proposing a novel hardware approximation method for the sigmoid function.  The method utilizes a combination of first- and second-degree polynomial functions to approximate the sigmoid curve, with the primary goal of minimizing approximation error while optimizing hardware resource usage.  This approach offers a practical solution for deploying efficient and accurate neural activation functions in FPGA-based systems, paving the way for improved performance in neuromorphic applications. 
\section{Related Work}

The sigmoid function is a nonlinear component widely used in machine learning as a neural activation function. Artificial neural networks (ANNs) implemented with neuromorphic hardware provide solutions for various applications where high data throughput, small size and less power consumption are required. Field-Programmable Gate Arrays (FPGAs) owns these essential features that are aligned with the requirements. However, due to the trade-off between resource limitations and numerical accuracy, implementing ANNs on FPGAs also poses difficulties, such as the computing of the neuron activation functions. 
Bosso, V.d.A et.al [1] \cite{p18}, the authors propose a systematic domain-partitioning strategy for sigmoid approximation that identifies optimal boundaries rather than relying on heuristic or arbitrary segments. This optimization resulted in a low mean absolute approximation error of $1.66\times{10}^{-3}$ by employing $0.04\%$ logic blocks and $3.21\%$ DSP blocks in a Ciclone V 5CGXFC7C7F23C8 FPGA Device, a marked improvement over previous methods. The architecture relaxes the traditional constraint of power-of-two operands, demonstrating that arbitrary operand values can be utilized without a systematic increase in hardware overhead. The work establishes a rigorous methodological framework by validating results across diverse FPGA architectures from multiple vendors and providing open-source VHDL implementations of all compared designs to ensure reproducibility. The study advocates for a standardized benchmarking process, utilizing consistent bit-widths, synthesis tools, and hardware targets to ensure equitable performance comparisons within the field.
Number of papers that have been published on the hardware implementation of the sigmoid function also demonstrate the ongoing interest in studying it. The sigmoid function is commonly implemented using piecewise approximation with linear or nonlinear segments, LUTs for direct output value storage, Taylor series expansion, and Coordinate Rotation Digital Computer (CORDIC) techniques. The study \cite{p2} uses CORDIC-based methods to iteratively compute trigonometric and hyperbolic functions with the goal of achieving low maximum approximation error. Nevertheless, this method requires a large amount of chip resources. Similarly, the authors in \cite{p3} used the Lagrange form and Taylor's theorem for approximation. To improve chip space efficiency, the paper also suggests reusing neuron circuitry in the approximation calculation.

\section{Proposed Methodology}
The sigmoid activation function plays a fundamental role in a wide range of artificial intelligence and machine learning (AI/ML) applications. It is extensively used in neural networks for tasks such as binary classification, probabilistic modeling, and gating mechanisms in recurrent architectures. One important aspect of this function is its ability to take an input of any value, whether positive or negative, and translate it into an output value that will only fall within the range of 0 to 1, which can be easily viewed as a probability. This feature enables it to be utilized within the output layers of classifications, logistic regression models, as well as in attention and control units within a system of deep learning. This mathematical definition of this function is presented in \eqref{sig_1}.
\begin{align}
    \sigma(x) = \frac{1}{1 + e^{-x}}
    \label{sig_1}
\end{align}

The direct hardware implementation of the exponential function involved in the sigmoid is computationally expensive. Many researchers have explored an alternative way that enables efficient implementation while preserving numerical accuracy. The sigmoid can be expressed in terms of tanh as given in \eqref{sig_2}.

\begin{align}
    \sigma(x) = \frac{1}{2}\left(1 + \tanh\left(\frac{x}{2}\right)\right)
     \label{sig_2}
\end{align}

From a design perspective, this identity reduces the process of computing the sigmoid function down to hyperbolic tangent (tanh) evaluation, followed by the operations of a scale and an add. In addition, the hyperbolic tangent (Tanh) can be calculated very efficiently through CORDIC-based architectures, which only require the use of shifting and addition~\cite{v1}. Therefore, this means that this new approach has significantly reduced the amount of computations needed for the evaluation of sigmoids and eliminated the requirement for both multipliers and exponential units.

The proposed method assumes that the input range of the sigmoid activation function will be limited to $\pm 1$. The majority of practical AI/ML systems, when developing a neural network model, normalize the input to the neural network to improve the convergence range, thus allowing for better numerical stability. With a limited input range, it is possible to optimise the design of the neural network model to focus on the most relevant region while not sacrificing the performance of the model as a whole.

Based on a sigmoid–tanh relationship, the corresponding range for the sigmoid inputs, $\pm 1$, is mapped to tanh, with only output values between $\pm0.5$. This narrow input range offers a significant advantage when implementing on hardware. The output from tanh will converge quicker with fewer iterations and less error when using iterative algorithms (like CORDIC). Thus, the overall latency, area, and power are reduced when using the sigmoids activation functions versus tanh.

The CORDIC algorithm allows for the efficient and hardware-compatible computation of all transcendental functions, including trigonometric and hyperbolic functions~\cite{an1,an2}. CORDIC was originally designed for navigation systems where real-time response is critical, but it is now particularly useful for modern artificial intelligence and machine learning (AI/ML) hardware accelerators that are based on field programmable gate arrays (FPGAs) and edge devices, this is due to the fact that it uses shift-and-add operations exclusively. Therefore, there is no requirement for either multipliers or complex exponential functions; consequently, CORDIC decreases the area, power, and complexity of designs.

In hyperbolic mode, the CORDIC algorithm can be configured to compute 
sinh and cosh. The hyperbolic tangent can then be obtained as the ratio of these two quantities. In the proposed approach, CORDIC is used to generate 
tanh within a limited and well-defined input range. As discussed earlier, by expressing the sigmoid function in terms of the hyperbolic tangent, the problem of computing the sigmoid reduces to evaluating $tanh(\frac{x}{2})$ followed by a simple scaling and offset operation. Since the input range of the sigmoid function is assumed to be limited to $\left[-1,1 \right]$, the corresponding input range for the CORDIC-based tanh computation becomes $[-0.5,0.5]$. This reduced range significantly improves the convergence behavior of the hyperbolic CORDIC algorithm and allows the number of iterations to be minimized without compromising accuracy.
\begin{figure}
    \centering
    \includegraphics[width=1\linewidth]{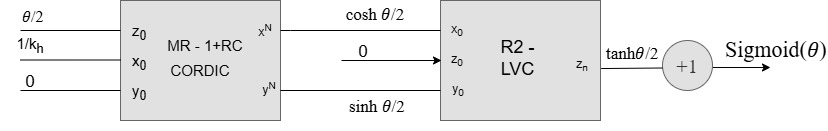}
    \caption{Proposed CORDIC-based methodology organized as a three-stage processing pipeline}
    \label{fig1}
\end{figure}
As shown in Fig.\ref{fig1}, the proposed sigmoid realization is organized as a three-stage processing pipeline. The first stage uses a mixed-radix HRC, which accepts initial input $(x_0=\frac{1}{K_h},y_0=0,z_0=Z_{in})$. In this stage, the CORDIC operates in hyperbolic rotation mode to iteratively compute the required hyperbolic sinh and cosh. The MR-HRC forms the computational core for generating the hyperbolic $sinh(Z_{in})$ and $cosh(Z_{in})$ using shift–add operations.

Outputs from the HR-CORDIC stage are sent to the second stage consisting of a hyperbolic vectoring block (LV-CORDIC). This stage refines the hyperbolic results and calculates the tanh-related value from the rotated vector by using vectoring mode operations. By using LV-CORDIC, the design efficiently calculates the ratio of hyperbolic components without explicitly including any divisions or multiplications, thus maintaining a DSP-free datapath.

Since there is a well-known relationship between the hyperbolic tangent and the sigmoid function, we can use the tanh output from an LV-CORDIC block to find a corresponding scaled and offset version of that output. The operation required to go from tanh to sigmoid is simply an addition operation, followed by scaling the resulting number by two constant values. Because this stage consists solely of addition and scaling by constant values, it does not have much impact on the overall hardware resource utilization.

The R2-CORDIC is a traditional way of obtaining rotation using angles and coordinate points~\cite{an3}. R2-CORDIC was originally created because of its easy implementation and offers good precision, but it has a slow convergence rate; thus, more iterations are needed to obtain very high precision results. The additional number of iterations means that there would be longer latencies and slower throughput. Therefore, to take advantage of both Radix-4 and Radix-2 CORDIC algorithms, a MR-HRC implementation combining Radix-4 and Radix-2 CORDIC for stable convergence is proposed.

As a result, the mixed-radix approach significantly reduces the total number of iterations required to achieve a given accuracy compared to a pure radix-2 implementation. This leads to lower computational latency and improved hardware efficiency, while preserving numerical stability. Hence, the proposed mixed-radix CORDIC architecture offers an effective trade-off between convergence speed and implementation complexity, making it well-suited for high-performance and resource-constrained hardware platforms. In the following section, we have discussed the proposed modified mixed-radix hyperbolic rotation CORDIC (MR-HRC) approach to derive hyperbolic sine and cosine.
\subsection {Mixed-Radix Hyperbolic Rotation Cordic(MR-HRC)}
In this section, we have discussed the proposed mixed-radix hyperbolic rotation CORDIC (HRC) algorithm, which combines radix-2 and radix-4 iterations to overcome the slow convergence of conventional radix-2 CORDIC. A radix-R hyperbolic rotation CORDIC (HRC) computes hyperbolic functions by applying a sequence of micro-rotations whose step sizes shrink geometrically with the iteration index. At iteration j, the elementary hyperbolic angle is typically chosen as follows:
\begin{align}
\alpha_j = \tanh^{-1}\!\left(d_jR^{-j}\right)
\label{angl}
\end{align}
The variable $d_j$ in \eqref{angl} indicates the digit selection function. In CORDIC, the convergence range means the set of input angles for which the iterative micro-rotations are guaranteed to reach the target angle without losing stability. The convergence range of the radix-R HRC CORDIC algorithm is defined in \eqref{conv_rng}.

\begin{equation}
|z_0| \le \sum_{i=1}^{N} \tanh^{-1}\!\left(d_{j,max}R^{-j}\right)
\label{conv_rng}
\end{equation}
where \(d_{j,\max}\) represents the maximum allowable digit magnitude and \(N\) is the total number of iterations. As discussed earlier, to compute the sigmoid function over the normalized  input range \(x \in [-1, 1]\), the identity $ \sigma(x) = \frac{1}{2}\left(1 + \tanh\left(\frac{x}{2}\right)\right) $ is employed. As a result, the required input range for the hyperbolic tangent function is reduced to $Z_0 \in \left[-\frac{1}{2},\,\frac{1}{2}\right]$. This restricted input range is particularly advantageous for hardware implementations, as it places the CORDIC core in a small-angle operating region, leading to improved numerical stability and reduced iteration requirements.

By substituting \(R=2\) and \(R=4\) into \eqref{conv_rng}, the convergence range of the radix-2 and radix-4 CORDIC can be defined as given in \eqref{r2} and \eqref{r4}, respectively.
\begin{equation}
|z_0| \le \sum_{j=1}^{N} \tanh^{-1}\!\left(2^{-j}\right)
\label{r2}
\end{equation}
\begin{equation}
|z_0| \le \sum_{j=1}^{N} \tanh^{-1}\!\left(2*4^{-j}\right)
\label{r4}
\end{equation}

The required input convergence range for the proposed sigmoid computation is limited to $\pm0.5$, as established earlier through the sigmoid–tanh relationship. Based on this requirement, the R2-HRC algorithm is selected for the initial stage, with the iteration index chosen to start with $j=2$. Using the convergence criteria defined in \eqref{r2}, the allowable input range for R2-HRC with the iteration index $j=2$ can be evaluated to approximately 0.5688, which is greater than the required input range of 0.5. Hence, starting the R2-HRC iterations with $j=2$ guarantees convergence for the required input angle. As a result, in the proposed architecture, the first stage performs R2-HRC iterations from $j=2$ to $j=9$. This stage ensures that the residual angle is brought down to a smaller range suitable for subsequent higher-radix rotation stages. 

After completing the radix-2 HRC rotation from $j=2$ to $j=9$, the residual angle is reduced to approximately $0.0061$. To continue refinement with a higher convergence speed, the next stage is initiated using R4-HRC. When the radix-4 computation starts from j=4, the corresponding admissible input range is 0.0104, which covers the residual angle produced by the R2-HRC. As a result, the transition from R2-HRC to R4-HRC satisfies the required convergence condition and does not generate an error. Therefore, the proposed design proceeds with the R4-HRC iteration in the subsequent stage.

A key limitation of conventional R4-HRC is that its scale factor is not constant; it varies with both the input angle and the digit selection function $d^j$. This input-dependent gain complicates hardware normalization and can introduce additional error if not compensated accurately. To avoid this issue, the proposed approach starts the R4-HRC stage at $j=4$. When the iteration is initiated with $j=4$, the rotation angle becomes sufficiently small such that the higher-order terms in the Taylor series expansion of the hyperbolic functions can be safely neglected without introducing any numerical error. Under this condition, the series effectively reduces to its dominant linear terms, and the associated gain converges to unity. As a result, the effective scale factor is approximately unity, so the rotation does not require an explicit scaling correction and does not introduce noticeable gain-related error. Therefore, initializing R4-HRC at j=4 enables an error-free and hardware-friendly implementation. 

In a radix-4 CORDIC algorithm, the digit selection function controls the amount of rotation in each iteration. Unlike radix-2 CORDIC, where the digit selection function is simple, and it depends only on the sign of the residual, radix-4 CORDIC uses a multi-valued digit set, $d^j \in \{-2,-1,0,+1,+2\}$, which allows fast convergence. In R4-HRC, the digit selection logic is more complex than that of radix-2 CORDIC. In radix-2, the digit is determined solely by the sign of the residual angle $Z^j$, resulting in a simple binary choice. In contrast, R4-HRC employs a multi-valued digit set, and the digit selection depends on both the sign and magnitude of the residual angle. The implementation of the complex digit selection function requires an additional comparator. 

In the proposed approach, the digit selection function for the R4-HRC is derived using the radix-4 SRT division. The only essential requirement imposed on the digit selection is that the magnitude of the residual angle must decrease after each iteration. The proposed digit selection ensures faster convergence compared to the conventional radix-2. In the proposed digit selection, the radix-4 SRT division is employed to derive the digit selection in each R4-HRC iteration. The scaled residual angle $4^jZ^j$ must be within a predefined interval bounded by a lower limit $(L^j\left[ \sigma^j\right])$ and an upper limit $(U^j\left[ \sigma^j\right])$ as defined in \eqref{lu}.
\begin{align}
    L^j[\sigma^j] = P^j[\sigma^j] - \frac{2}{3} P^j[1] \nonumber \\ 
    U^j[\sigma^j] = P^j[\sigma^j] + \frac{2}{3} P^j[1]
    \label{lu}
\end{align}
where, $ P^j[\sigma^j]=4^j\tanh^{-1}\!\left(\sigma^j4^{-j}\right)$. To determine the selection criteria for the digit $\sigma^j=2$, overlapping between two intervals $I^1: (L^j[1], U^j[2])$ and $I^2: (L^j[2], U^j[2])$ is evaluated. In the proposed approach, any value within the overlapping region between the intervals $I^1$ and $I^2$ can be chosen to select the digit $\sigma^j$. The values selected for $\sigma^j$ are carefully chosen to ensure that the required comparison logic for digit selection can be implemented efficiently in hardware. These values are represented in binary using 4 binary bits for simplicity and to minimize the hardware complexity as shown in \eqref{digit_sel}.
\begin{equation}
\sigma^j = 
\begin{cases} 
2 & \text{for } 4^jZ^j \ge 1.5, \\
1 & \text{for } 1.5 > 4^jZ^j \ge 0.5, \\
0 & \text{for } 0.5 > 4^jZ^j \ge -0.5, \\
-1 & \text{for } -0.5 > 4^jZ^j \ge -1.5, \\
-2 & \text{for } -1.5 > 4^jZ^j.
\end{cases}
\label{digit_sel}
\end{equation}
This approach significantly reduces the required comparison logic, allowing for a more hardware-efficient implementation. Next, we have explained the architecture to implement the same.
\subsection{Radix-2 LVC}

In the proposed algorithm, the R2-LVC algorithm is used to calculate the hyperbolic tangent (\(\tanh\)) from the sine (\(\sinh\)) and cosine (\(\cosh\)) values, which are computed using the MR-HRC CORDIC algorithm in the previous stage. The objective of the R2-LVC is to compute the $ \tanh(x) = \frac{\sinh(x)}{\cosh(x)}$ by performing vectoring mode operations where the R2-LVC is initialized with $\left(X^0=\cosh\, Y^0=\sinh\, Z^0=0 \right)$.
After the convergence, the value of $\tanh\ $ can be calculated from the converged value of the $Z^N$. The rotation equations for the R2-LVC are as follows:
\begin{align}
    x^{j+1} &= x^j \nonumber \\
    y^{j+1} &= y^j - d^j \cdot x^j \cdot 2^{-j}. \nonumber \\
    z^{j+1} &= z^j + d^j \cdot (2^{-j})
    \label{r2_lvc}
\end{align}

These equations are iteratively computed until the variable \(y^j\) is sufficiently small, achieving convergence. The final value of \(\tanh(x)\) is directly obtained from the variable \(z^j\).

The convergence range of the R2-LVC algorithm is defined as $\left|\frac{Y_0}{X_0}\right| \leq 2$. This condition ensures that the CORDIC algorithm can successfully converge to the desired result. If the ratio exceeds 2, the algorithm may fail to converge correctly. In the proposed approach, the input range for the MR-HRC algorithm is constrained to \(\pm 0.5\). The maximum value of the tangent ratio is $\frac{Y_0}{X_0} = \tanh(0.5) \approx 0.52$, which is well within the convergence range of the R2-LVC algorithm. Therefore, the input range of \(\pm 0.5\) is suitable for the CORDIC algorithm, ensuring stable and accurate convergence without exceeding the required convergence bounds.

\subsection{Architecture of the proposed approach}
\begin{figure}
    \centering
    \includegraphics[width=0.8\linewidth]{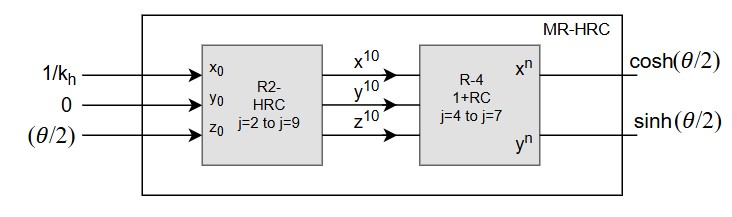}
    \caption{The architecture of proposed MR-HRC algorithm}
    \label{fig_mrc}
\end{figure}

\begin{figure}
    \centering
    \includegraphics[width=0.8\linewidth]{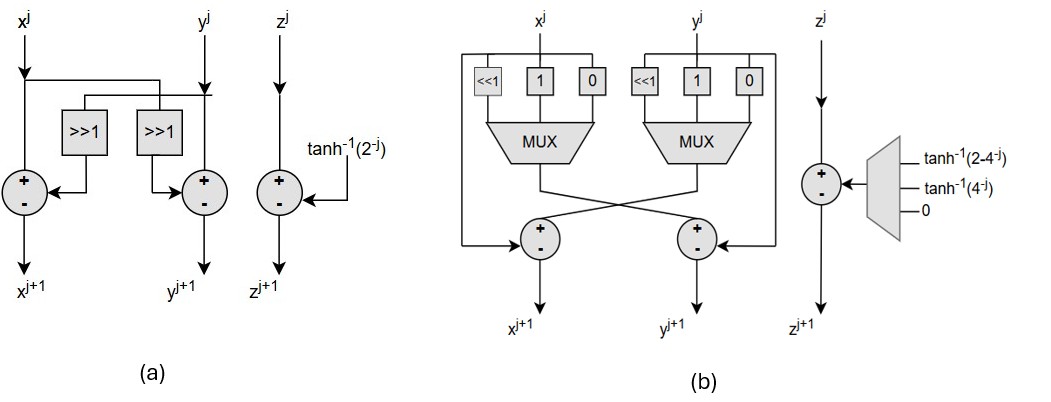}
    \caption{(a)Architecture of R2-HRC algorithm(b)Architecture of R4-HRC algorithm}
    \label{r2hrc}
\end{figure}

The architecture of the proposed approach is structured into two main stages for efficiently computing the hyperbolic functions sinh, cosh, and tanh. The 16-bit datawidth has chosen for hardware realization as it is sufficient for most of the real-time applications.  The first stage of the architecture involves a mixed-radix hyperbolic rotation CORDIC (HRC) algorithm. This stage is responsible for computing the hyperbolic sine (sinh) and hyperbolic cosine (cosh) functions. The mixed-radix approach combines radix-2 and radix-4 CORDIC iterations. Initially, the R2-HRC is applied to reduce the residual angle and get an approximate value of sinh(x) and cosh(x). The algorithm performs radix-2 CORDIC operations for iterations starting from j=2 to 9. The algorithm performs radix-4 CORDIC operations for iterations from j=4 to 7. Later, the R4-HRC further refines the results, providing more precise values of sinh(x) and cosh with faster convergence compared to using radix-2 alone. Fig.\ref{fig_mrc} shows the overall architecture of the MR-HRC algorithm, whereas Fig. \ref{r2hrc}(a) and Fig. \ref{r2hrc}(b) show R2-CORDIC and R4-CORDIC architectures, respectively.
 
\begin{figure}
    \centering
    \includegraphics[width=0.65\linewidth]{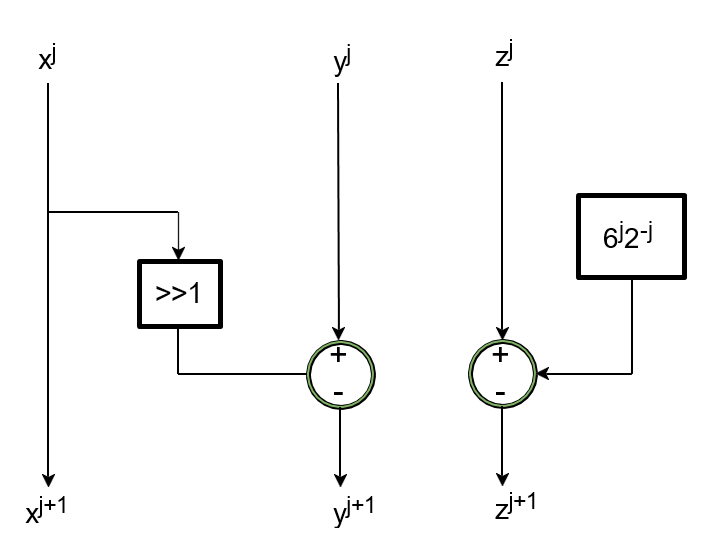}
    \caption{The architecture of the R2-LVC algorithm with adder}
    \label{lvc}
\end{figure}

In the proposed approach, we have implemented a pipelined architecture, where each iteration of the CORDIC algorithm has dedicated hardware, and pipeline registers are used to separate each iteration. This architecture ensures that the operations within each iteration are performed in parallel, which optimizes throughput and reduces the overall latency.
Due to the pipeline architecture, the adder is the only component that determines the critical path of the R2-HRC algorithm. In the R4-HRC algorithm, the digit selection function has five different values. As a result, the critical path involves two key components: the multiplexer and the adder. The architecture of the R2-LVC algorithm is quite simple, and it is illustrated in Fig. \ref{lvc}. The critical datapath of the R2-LVC algorithm has only an adder.

\section{Simulation Results}
The proposed architecture is implemented using a 16-bit data width, which provides sufficient numerical precision for most real-time applications while keeping the hardware cost low. This choice offers a practical balance between accuracy and resource efficiency, making the design suitable for time-critical signal processing and AI/ML workloads.

To evaluate its effectiveness, the proposed architecture is implemented on Xilinx Virtex-7 series FPGAs and compared with several recent state-of-the-art approaches. The comparison is carried out using standard metrics, including hardware utilization and mean absolute error. Hardware utilization reflects the efficiency of the design in terms of FPGA resources, and MAE captures the numerical accuracy of the computed results. The simulation results of the proposed architecture are compared to existing implementations using the discussed performance metrics.

The proposed architecture is synthesized using Xilinx Vivado and implemented on a Virtex-7 series FPGA to ensure a fair comparison with the previous approaches. Table \ref{tab_hw} illustrates a comparison of hardware utilization for different sigmoid implementations, where DSP usage has been converted into an equivalent logic cost for a fair comparison. In this table,
\begin{table}
\centering
\caption{Comparison of LUT, DSP, and total hardware utilization for FPGA-Based sigmoid implementations}
\begin{tabular}{|c|c|c|c|c|}
\hline
Work     & LUTs & DSPs & DSP-Equivalent LUTs & Total LUTs \\ \hline
\cite{p17} & 202  & 4    & 1540                 & 1742                   \\ \hline
\cite{p13} & 168  & 5    & 1925                 & 2093                   \\ \hline
\cite{p14} & 86   & 3    & 1155                 & 1241                   \\ \hline
\cite{p15} & 114  & 0    & 0                    & 114                    \\ \hline
\cite{p16} & 282  & 8    & 3080                 & 3362                   \\ \hline
\cite{p18} & 80   & 4    & 1540                 & 1620                   \\ \hline
Proposed & 835  & 0    & 0                    & 835                    \\ \hline
\end{tabular}

\label{tab_hw}
\end{table}

 one DSP block is treated as equivalent to 385 logic slices, and the total logic equivalent is computed by combining LUT usage with the DSP-equivalent LUTs. From the table, it can be observed that several existing approaches rely heavily on DSP blocks. Although reported hardware utilization may appear lower, the effective hardware utilization increases significantly once DSP usage is taken into account. For example, the approaches reported in \cite{p16} and \cite{p13} show total logic equivalents of 3362 and 2093 slices, respectively, primarily due to high DSP utilization. Even designs with fewer DSPs, such as \cite{p17} and \cite{p18}, still have a substantial hardware utilization. In contrast, the proposed approach is implemented without DSP blocks, relying entirely on LUT-based shift-and-add operations. As a result, the total logic equivalent of the proposed design is only 835 slices, which is significantly lower than most of the compared methods. When compared with DSP-intensive implementations such as \cite{p16} and \cite{p13}, the proposed architecture achieves nearly $40\%$ improvement in overall hardware utilization, highlighting its efficiency.

\begin{table}
\centering
\caption{Comparison of mean absolute error (MAE) for different sigmoid computation methods}
\begin{tabular}{|c|c|}
\hline
Work & Average Absolute Error \\
\hline
\cite{p17} & $1.71 \times 10^{-3}$ \\ 
\cite{p13} & $1.07 \times 10^{-3}$ \\
\cite{p14} & $4.25 \times 10^{-3}$ \\
\cite{p15} & $5.90 \times 10^{-3}$ \\
\cite{p16} & $2.44 \times 10^{-3}$ \\
\cite{p18} & $1.66 \times 10^{-3}$ \\
\textbf{Proposed} & $\mathbf{4.23 \times 10^{-4}}$ \\
\hline
\end{tabular}
\label{mae}
\end{table}

Table \ref{mae} presents a comparative analysis of the mean absolute error achieved by various existing approaches and the proposed method. The mean absolute error is a widely used metric to quantify the numerical accuracy of function approximation algorithms, as it reflects the average deviation between the computed output and the ideal reference value over a large set of input samples. From the Table \ref{mae}, it can be observed that several existing approaches have mean absolute errors in the range of $10^{-3}$. The approached reported in \cite{p14} and \cite{p15} show relatively higher errors of $4.25 \times 10^{-3}$ and $5.90 \times 10^{-3}$, respectively, indicating limited approximation accuracy. Approaches such as \cite{p13} and \cite{p17} still maintain errors above $10^{-3}$. The approach presented in \cite{p18} achieves a moderate improvement; however, its error remains significantly higher than that of the proposed design. In contrast, the proposed approach achieves a mean absolute error of only $4.23 \times 10^{-4}$, which is substantially lower than all the compared methods. This represents a significant improvement in numerical accuracy, demonstrating the effectiveness of the proposed mixed-radix CORDIC-based formulation and the careful selection of iteration ranges.
\section{Conclusion}
We have presented a hardware-efficient FPGA implementation of the sigmoid activation function based on a mixed-radix hyperbolic rotation CORDIC architecture in this paper. We have used the mathematical relationship between the sigmoid and hyperbolic tangent functions; the proposed approach effectively converts the expensive sigmoid computation into a sequence of CORDIC-based hyperbolic rotation and division operations followed by simple scaling and offset adjustments. We have normalized the input range to $\pm 1$, which further reduces the convergence range of the MR-HRC algorithm.
The main objective of this work is to design a modified mixed-radix hyperbolic rotation CORDIC (MR-HRC) algorithm that combines radix-2 and radix-4 iterations. The radix-2 stage guarantees the initial convergence for the required input range. The input $X^0$ of the algorithm is initialised with $\frac{1}{K_h}$ so that the scale factor generated by radix-2 CORDIC can be compensated without any additional hardware. Later, the radix-4 stage makes convergence faster by producing two bits of result in each iteration. The subsequent radix-2 linear vectoring CORDIC (R2-LVC) stage efficiently computes the hyperbolic tangent using only shift-and-add operations, resulting in a DSP-free datapath. The proposed architecture is fully pipelined and implemented using a 16-bit fixed-point representation, making it well-suited for real-time and resource-constrained applications. Implementation on an Xilinx Virtex-7 FPGA demonstrates that the design achieves a total logic equivalent of only 835 slices with zero DSP usage. In addition, the achieved mean absolute error of $4.23 \times 10^{-4}$ confirms the high numerical accuracy of the proposed method. The proposed MR-HRC formulation can be extended to other nonlinear functions and integrated into larger neuromorphic systems, making it a low-power and high-performance AI hardware.
\bibliographystyle{splncs04}
\bibliography{mybibliography}
\end{document}